\typein[\format]{landscape or portrait (l/p)?}
\if l\format
\documentstyle[12pt,2col]{article}
    \twocolumn
    
    \evensidemargin= \oddsidemargin
\else
\documentstyle[12pt]{article}
\fi

\newcommand{\mysection}[1]{\setcounter{equation}{0}\section{#1}}

\begin{document}
\setlength{\baselineskip}{0.20in}
\def\thefootnote{\fnsymbol{footnote}}

\newcommand{\nc}{\newcommand}
\nc{\Eq}{\begin{equation}}   \nc{\Endl}[1]{\label{#1} \end{equation}}
\nc{\End}{\end{equation}}    \nc{\Eqa}{\begin{eqnarray}}
\nc{\Endla}[1]{\label{#1} \end{eqnarray}}
\nc{\Enda}{\end{eqnarray}}   \nc{\Eqr}[1]{(\ref{#1})}

\nc{\Fp}{\tilde{\phi}}   \nc{\FP}{\tilde{\Phi}}
\nc{\FA}{\tilde{A}}      \nc{\AMC}{A^{\mu}_c}
\nc{\Am}{{\cal A} }      \nc{\iibar}{\mbox{$I$--$\bar{I}\,$}}
\nc{\ssbar}{\mbox{$S$--$\bar{S}\,$}}

\nc{\ib}[3]{        {\em ibid. }{\bf #1} (19#2) #3}
\nc{\np}[3]{        {\em Nucl. Phys. }{\bf #1} (19#2) #3}
\nc{\pl}[3]{        {\em Phys. Lett. }{\bf #1} (19#2) #3}
\nc{\pr}[3]{        {\em Phys. Rev.  }{\bf #1} (19#2) #3}
\nc{\prep}[3]{      {\em Phys. Rep.  }{\bf #1} (19#2) #3}
\nc{\prl}[3]{       {\em Phys. Rev. Lett. }{\bf #1} (19#2) #3}

\begin{titlepage}

{\hbox to\hsize{October 1993 \hfill HUTP-93-A035}}
{\hbox to\hsize{ \hfill JHU-TIPAC-930028}}
{\hbox to\hsize{ \hfill hep-ph/9311291}}

\begin{center}
\vskip .5 in
\if l\format
{\Large \bf Spacetime Symmetries and Semiclassical Amplitudes}
\else
{\large \bf Spacetime Symmetries and Semiclassical Amplitudes}
\fi
\vskip .4 in

\begin{tabular}{cc}
\begin{tabular}{c}
{\bf Thomas M. Gould\footnotemark[1]}\\[.05in]
{\it Department of Physics and Astronomy}\\
{\it The Johns Hopkins University}\\
{\it Baltimore MD 21218 }\\[.15in]
\end{tabular}
&
\begin{tabular}{c}
{\bf Stephen D.H. Hsu}\footnotemark[2]\\[.05in]
{\it Lyman Laboratory of Physics} \\
{\it Harvard University} \\
{\it Cambridge  MA 02138} \\[.15in]
\end{tabular}
\end{tabular}

\end{center}

\footnotetext[1]{Email:\tt gould@fermi.pha.jhu.edu}
\footnotetext[2]{SSC Fellow. Email: \tt Hsu@HUHEPL.bitnet,
Hsu@HSUNEXT.Harvard.edu}

\begin{abstract}
\smallskip
We examine the spacetime symmetries of forward $2 \rightarrow 2$
scattering.  These symmetries have non-trivial consequences for
any class of configurations which might dominate the amplitude
in the semiclassical approximation.  We derive some dynamical
results regarding the stability of configurations which arise
solely from reflection symmetry and positivity of the (Euclidean)
path integral action.  We show that in the case of on-shell
scattering, any semiclassical configuration must have $O(3)$
symmetric asymptotic behavior.  We argue that non--$O(3)$
symmetric configurations are irrelevant to the semiclassical
exponent.

\end{abstract}

\end{titlepage}

\renewcommand{\thepage}{\arabic{page}}
\setcounter{page}{1}
\mysection{Forward scattering in semiclassical approximation}

Recent interest in baryon number violation at high energies~\cite{BV}
has led to attempts to compute nonperturbative
contributions to the imaginary part of $2\rightarrow 2$
forward scattering amplitudes derived from correlators such as
\if l\format
\Eqa
\lefteqn{
\langle ~\Fp (k_1) ~\Fp(k_2) ~\Fp^* (k_1) ~\Fp^* (k_2) ~\rangle \: = \: }
& & \label{fsa} \\
& & \hspace{2.5cm} N \int D\phi ~DA ~
\Fp (k_1) \ldots \Fp^* (k_2) ~e^{i S\left[\Phi,A\right]} ~.\nonumber
\footnotemark[1]
\Enda
\footnotetext[1]{
The integral in (\ref{fsa}) should be understood as continued
at least infinitesimally ($t \rightarrow t e^{-i \theta}$) to
Euclidean space to ensure convergence.}
\else
\Eq
\langle ~\Fp (k_1) ~\Fp(k_2) ~\Fp^* (k_1) ~\Fp^* (k_2) ~\rangle \: = \:
N \int D\phi ~DA ~
\Fp (k_1) \ldots \Fp^* (k_2) ~e^{i S\left[\Phi,A\right]} ~. \nonumber
\footnote{The integral in (\ref{fsa}) should be understood as continued
at least infinitesimally ($t \rightarrow t e^{-i \theta}$) to Euclidean
space to ensure convergence.}
\Endl{fsa}
\fi
Here, $\Fp (k)$ is meant to represent the Fourier transform of either
the Higgs scalar ($\Phi $) or
the vector boson ($A^\mu $) fields of the Standard Model.
(We will also study the case in which the fields are fermionic below.)
By computing the imaginary part of this amplitude,
one obtains the total cross section $\sum_{n} \sigma(2 \rightarrow n)$.
If the calculation of (\ref{fsa}) can be restricted to
the anomalous baryon number violating sector,
then the result is the total cross section for baryon number violation
from a two particle initial state.
Of course, there are many difficulties associated with this program:
the problem of how to project (\ref{fsa}) onto the anomalous sector has
not been satisfactorily addressed except at low energies~\cite{ZEROMODES},
and it is not clear that (\ref{fsa}) is dominated by any single semiclassical
configuration at high energies, $E\simeq E_{sph}\sim M_w/g^2$.
(Although see \cite{MMY} for some plausible arguments that this may be
the case.)
Furthermore, the variational equations which must be solved to find
a saddlepoint for the path integral expression for (\ref{fsa}) are of
a highly nontrivial integral-differential type and will in general
lead to a complex saddlepoint~\cite{MMY}.

In this note,
we will {\it assume} that a semiclassical approximation to (\ref{fsa}) exists.
Furthermore,
we will initially be interested in the case that the semiclassical behavior
is determined by a {\it single} classical ``master field'' configuration
(up to translations).
Then,
we will use the spacetime symmetries of the two particle forward
scattering amplitude to constrain such a configuration.
In the semiclassical approximation, (\ref{fsa})
becomes
\Eq
\langle ~\Fp_c (k_1) ~\Fp_c (k_2) ~\Fp_c^* (k_1) ~\Fp_c^* (k_2) ~\rangle ~,
\Endl{sca}
where $\Fp_c (k)$ is the Fourier transform of the corresponding classical field
configuration $\Phi_c (x)$
or $\AMC (x)$. Symmetry properties of the amplitude then translate to symmetry
properties of the
classical configurations. In the alternate case of a {\it family} of degenerate
saddlepoints, the
family will be related by the symmetries of the amplitude.

Consider the $2 \rightarrow 2$ scattering amplitude
$\Am( k_a, k_b, k_c, k_d) = \Am(s,t,u)$, described by the Mandelstam
variables, $s = (k_a + k_b)^2, t = (k_a - k_c)^2$ and $u = (k_a - k_d)^2$.
(See figure 1.)
In the forward case, we choose without loss of generality,
$k_a = k_c$ and $k_b = k_d$, so that $t = 0$.
In the center of mass (CM) frame,
we can choose the particles to be collinear along the \mbox{$z$-axis}
so that
$s = (k_a^0 + k_b^0)^2 \equiv 4 E^2$ and $u = (k_a^z - k_b^z)^2 \equiv 4
(k^z)^2$.
Thus,
forward scattering amplitudes in the CM frame have certain manifest
symmetries:
$\Am$ depends only on the total energy $E$ squared and the $z$ component
of the momentum $k$ squared.
This leads to the following symmetries in momentum space:
an $O(2)$ symmetry associated with rotations about the \mbox{$z$-axis},
and
discrete symmetries associated with changing the signs of $E$ or $k^z$.
These symmetries, combined with the properties of the Fourier transform,
imply spacetime symmetries for the fields $\Phi_c (x)$ and $\AMC (x)$.
Under $O(2)$ rotations, or reflections through the
$t=0$ or $z=0$ planes (labelled by operators $T$ and $Z$
respectively),
the complex field $\Phi_c (x)$ can acquire at most a phase, while the real
field $\AMC (x)$ can acquire at most a sign.
(In the case of vector fields,
the above results are valid provided we take the polarization
$\epsilon^{\mu}$ of the vector fields to be orthogonal to the
$t-z$ CM collision plane. Then, there are no additional momentum
dependent Lorentz invariants that can be formed, and the
Mandelstam variables continue to fully parametrize the process.)

\if l\format
\vskip -0.8cm
\begin{picture}(300,200)(70,0)
\else
\vskip -0.8cm
\begin{picture}(300,200)
\fi
\put(150,100){\thicklines\circle{100}}
\put(135,115){\thicklines\line(-1,1){25}}
    \put(115,135){\thicklines\vector(1,-1){0}} \put(95,145){$k_a$}
\put(135,85){\thicklines\line(-1,-1){25}}
    \put(115,65){\thicklines\vector(1,1){0}} \put(100,50){$k_b$}
\put(165,85){\thicklines\vector(1,-1){25}} \put(195,50){$k_c$}
\put(165,115){\thicklines\vector(1,1){25}} \put(195,145){$k_d$}
\put(250,80){\fbox{
\shortstack{
$s = (k_a + k_b)^2$ \\$t = (k_a - k_c)^2$ \\$u = (k_a - k_d)^2$}
}}
\put(200,20){Fig. 1}
\end{picture}

Note that the symmetries listed above are all rotations or special
cases thereof.
Rotations are all that remain of the Lorentz group
when we choose the CM frame.
Of course,
the classical solutions $\phi_c (x)$ should remain solutions under
translation as well.

Let us quickly rule out the possibility of phases or minus signs
in the symmetry transformations of the classical fields.
Suppose that under $T$ or $Z$ reflection one of the
fields acquires a phase or sign change.
By continuity,
this requires a zero of the field on a three dimensional hypersurface
extending to infinity.
In the case of the Standard Model Higgs field,
this is precisely an infinite domain wall (or domain hypersurface),
and hence leads to infinite action for the configuration.
Similarly, a nontrivial phase under rotation implies an infinite two
dimensional surface of zeros.
Now,
the problem of extremizing the path integral in (\ref{fsa})
does not simply require extremizing the Standard Model action,
but rather depends crucially on the prefactor of Fourier transforms.
(When energies are of order $E_{sph}\sim M_w/g^2$,
corrections coming from
the prefactor cannot be neglected relative to the classical action.)
However,
in a massive theory, the Fourier transforms $\Fp (k)$ are well behaved,
and hence a semiclassical configuration with infinite action is still
arbitrarily suppressed in the path integral.
Therefore,
we can ignore the possibility of phases in the Higgs symmetry transformation,
and conclude that $\Phi_c (x)$ is exactly invariant under
$T,Z$ and $O(2)$ rotations.
The vector fields $\AMC (x)$ can acquire a  minus sign under $T$ or $Z$,
as in general a zero of the vector field does  not imply a nonzero action
density.
However,
since we have demonstrated that the
$\Phi_c (x)$ field is exactly symmetric under $T$ and $Z$,
and the equations of motion coupling $\Phi_c (x)$ and $\AMC (x)$
depend on the sign of $\AMC (x)$,
we can conclude that an extremal $\AMC (x)$ configuration will be symmetric.

\mysection{Reflection symmetry and \mbox{dynamics}}

Configurations with exact reflection symmetry can be shown to have an
interesting dynamical property
if the action $S\, [ \Phi, A^{\mu} ]$ is positive definite.
This condition is of course satisfied by any
Euclidean action,
and in particular also for certain choices of constraint such as
$\delta S = \int d^4x ~(F^2)^2$,
so the result can be applied to the modified actions relevant to
constrained solutions such as the instanton of the electroweak theory.
The property we will demonstrate is an instability in the direction of symmetry
(i.e.-- in the $z$ or $t$ directions).
We do so by modifying a proof due to
Aharonov, Casher, Coleman and Nussinov~\cite{ACCN}
which shows attraction at all distances between
\mbox{'tHooft}--Polyakov monopoles
and anti-monopoles.
The result implies that it is highly unlikely that a configuration with exact
reflection symmetry can
extremize a positive definite action (i.e.-- satisfy the equations
of motion), and that,
in the event that it does,
it must have a negative mode of a higher fluctuation operator such
as $\delta S / \delta \phi(x) \delta \phi(y)$.

\if p\format
\vskip -0.5cm
\begin{picture}(300,200)(-70,0)
\put(50,100){\vector(1,0){200}}\put(260,96){$t$}
\put(150,35){\line(0,1){130}}
\multiput(140,35)(0,20){7}{\line(0,1){10}}
\multiput(160,35)(0,20){7}{\line(0,1){10}}
\put(140,30){\vector(1,0){20}\vector(-1,0){20}}
\put(80,100){\circle{40}\circle*{2}}
\put(220,100){\circle{40}\circle*{2}}
\put(200,100){\line(1,1){20}}
\put(202,92){\line(1,1){26}}
\put(206,86){\line(1,1){28}}
\put(212,82){\line(1,1){26}}
\put(220,80){\line(1,1){20}}
\put(146,17){$\Delta $}
\put(135,0){Fig. 2}
\end{picture}
\vskip 0.5cm
\fi

The proof is as follows.
Consider a reflection symmetric configuration $\phi_c (x)$.
Now remove from the configuration an infinitesimally thin hyper-slice,
$\Delta$,
symmetric about the hypersurface of symmetry
(i.e.-- $t=0$ or $z=0$, see figure 2).
The symmetry condition implies that what remains
can be patched together to form a new configuration $\phi_c^\prime (x)$
in $R^4$ which
is continuous (although not necessarily in its derivatives), and has finite
action if $S$
does not contain terms higher than second order in derivatives.
Positivity of the action now guarantees that the new configuration satisfies
\Eq
S\left[\,\phi_c\,\right] ~\geq~ S\left[\,\phi_c^\prime\,\right] ~,
\End
with equality holding only in the exceptional case of exactly vanishing action
density on the hyper-slice
$\Delta$.
If there is nonzero action density in $\Delta$,
then $\phi_c (x)$ cannot be a solution to the equations of motion,
and in any case there must still be an instability of
the configuration to perturbations which move the mirror symmetric
configurations on either side of the planes $z,t =0$ closer together.

\if l\format
\vskip -0.5cm
\begin{picture}(300,200)(10,0)
\put(50,100){\vector(1,0){200}}\put(260,96){$t$}
\put(150,35){\line(0,1){130}}
\multiput(140,35)(0,20){7}{\line(0,1){10}}
\multiput(160,35)(0,20){7}{\line(0,1){10}}
\put(140,30){\vector(1,0){20}\vector(-1,0){20}}
\put(80,100){\circle{40}\circle*{2}}
\put(220,100){\circle{40}\circle*{2}}
\put(200,100){\line(1,1){20}}
\put(202,92){\line(1,1){26}}
\put(206,86){\line(1,1){28}}
\put(212,82){\line(1,1){26}}
\put(220,80){\line(1,1){20}}
\put(146,17){$\Delta $}
\put(135,0){Fig. 2}
\end{picture}
\vskip 0.5cm
\fi

As simple applications of this result,
we examine some of the configurations considered by various authors
as approximate saddlepoints to the $2\rightarrow 2$ scattering amplitude
in the Standard Model:
\if l\format \newpage \fi

1. {\em The Instanton Anti-instanton Pair} \\
Consider a field configuration consisting of the superposition of
an instanton and an antiinstanton  at some separation $R$.
If the relative orientations of the pair are in the ``attractive channel''
(i.e.-- the configuration is reflection symmetric),
then the above arguments prove that the action decreases monotonically
as the pair are brought together,
finally vanishing when the pair overlap exactly.
Note that this is a nonperturbative (albeit purely classical) result,
and applies also to the constrained action of an electroweak
instanton-antiinstanton pair.

2. {\em The Valley Configuration} \\
Our result also shows that for a positive definite choice of constraint,
the valley action~\cite{KR} decreases monotonically to zero for reflection
symmetric valleys.
This is consistent with the results of an explicit construction of
a valley in an {\it ad hoc} model of the electroweak theory~\cite{KR}.
The result however applies more generally to any valley configuration
in the full electroweak theory.
Although a valley in electroweak theory has not yet been explicitly
constructed, we have obtained non-trivial information about it.

3. {\em Klinkhamer's New Instanton $I^{\bf *}$} \\
Our result also has implications for the new solutions Klinkhamer
has found in the electroweak theory~\cite{KLI}.
The first solution $I^*$ is a new time-dependent Euclidean solution
to constrained electroweak equations of motion.
The solution resembles a bound state comprised of an instanton
and anti-instanton pair (\iibar ).
It can be thought of as a slice of a five dimensional constrained
instanton whose existence is implied by $\pi_4 [ SU(2)] = Z_2$.
In other words,
$I^*$ is the sphaleron of the Witten anomaly in five dimensions.
Klinkhamer has claimed that $I^*$ should dominate the amplitude
(\ref{fsa}) at high energies.
This claim is not obviously justified.
Although $I^*$ solves the equations of motion,
it is not clear at all that it extremizes the {\it product} of
the initial state factor and the exponential of the action.

The second solution $S^*$ is a static saddlepoint of the energy over
configuration space, which resembles a bound state comprised of
a sphaleron and anti-sphaleron pair (\ssbar ).
It plays the role of the original sphaleron $S$ in setting the height
of an energy barrier in configuration space,
but now for a noncontractible loop related to the Witten anomaly.

Since the action (energy) density is positive definite,
our result implies that the solutions  $I^*$ ($S^*$) cannot be mirror
symmetric, in either the $z$ or $t$ planes.
Indeed,
Klinkhamer's explicit solutions $I^*$ and $S^*$ exhibit relative
isospin rotations between their  \iibar and \ssbar constituents,
respectively.
This relative rotation allows $I^*$ ($S^*$) to extremize the action (energy).
However,
it also means that $I^*$ will lead to semiclassical scattering amplitudes
(\ref{fsa}) which in momentum space depend on {\it odd} powers the energy
$E$ and momentum $k_z$ (not just $E^2$ \mbox{and $k_z^2$).}

If $I^*$ is to lead to reasonable Lorentz invariant scattering amplitudes,
we must sum over discrete symmetry transformation of the configuration,
like integrating over translations of a typical soliton or instanton.
The forward scattering amplitude is then given by a sum of amplitudes,
each dominated by a reflection of $I^*$ through the $t$ or $z$ planes.
Indeed by writing,
\if l\format
\Eqa
\lefteqn{
\langle ~\Fp (k_1) ~\Fp(k_2) ~\Fp^* (k_1) ~\Fp^* (k_2)~
\rangle \:\simeq\: } & &\\
& & \langle ~\Fp_c (k_1) ~\Fp_c (k_2) ~\Fp_c^* (k_1) ~\Fp_c^* (k_2)~
\rangle~\rule[-3mm]{0.2mm}{8mm}_{\: I^*} \; + \; \nonumber \\
& & \hspace{1.0cm}
\langle ~\Fp_c (k_1) ~\Fp_c (k_2) ~\Fp_c^* (k_1) ~\Fp_c^* (k_2)
{}~\rangle~\rule[-3mm]{0.2mm}{8mm}_{\: T(I^*)} \; + \; \ldots ~,
\nonumber
\Enda
\else
\Eqa
\lefteqn{
\langle ~\Fp (k_1) ~\Fp(k_2) ~\Fp^* (k_1) ~\Fp^* (k_2)
{}~\rangle \:\simeq\:}  & & \\
& & \langle ~\Fp_c (k_1) ~\Fp_c (k_2) ~\Fp_c^* (k_1) ~\Fp_c^* (k_2)~
\rangle~\rule[-3mm]{0.2mm}{8mm}_{\: I^*} \; + \;
\langle ~\Fp_c (k_1) ~\Fp_c (k_2) ~\Fp_c^* (k_1) ~\Fp_c^* (k_2)~
\rangle~\rule[-3mm]{0.2mm}{8mm}_{\: T(I^*)}  \nonumber \\
& + & \langle ~\Fp_c (k_1) ~\Fp_c (k_2) ~\Fp_c^* (k_1) ~\Fp_c^* (k_2)~
\rangle~\rule[-3mm]{0.2mm}{8mm}_{\: Z(I^*)} \; + \;
\langle ~\Fp_c (k_1) ~\Fp_c (k_2) ~\Fp_c^* (k_1) ~\Fp_c^* (k_2)~
\rangle~\rule[-3mm]{0.2mm}{8mm}_{\: TZ(I^*)} ~, \nonumber
\Enda
\fi
we obtain an amplitude with the necessary discrete symmetries.

The $I^*$ example is instructive.
Given a configuration with {\it less} than
the full $O(2) \times T \times Z$ symmetry,
one can obtain an amplitude with the proper symmetry
by summing over amplitudes, each dominated by a member of the class
of degenerate configurations which are related by symmetry.
There are two possibilities in the extremization of (\ref{fsa}): either
\begin{enumerate}
\item there is a unique saddlepoint which exhibits exact
$O(2) \times T \times Z$ symmetry, or
\item there are classes of
degenerate configurations related by these symmetries.
\end{enumerate}

Finally, we should note a limitation of the preceding arguments.
While they apply to saddlepoints of the Euclidean action $S$,
they do not apply to saddlepoints of the Euclidean version of
(\ref{fsa}).
This formally involves the extremization of an ``effective'' action
\Eq
S_{eff}\left[\,\phi\,\right]\: = \: S\left[\,\phi\,\right]  \, - \,
\ln\left[ ~\Fp (k_1) ~\Fp(k_2) ~\Fp^* (k_1) ~\Fp^* (k_2) ~\right] ~,
\Endl{seff}
which takes into account the initial state factors~\cite{MMY}.
However, the initial state term is non-local
and cannot in general be
written as the integral of a (four) volume density.
So, the arguments given above unfortunately cannot be applied.
While reflection symmetric configurations do not in general
extremize $S$, they can however extremize $S_{eff}$,
as for example a widely separated \iibar pair
in the case of scattering at very low energies.

\mysection{$O(3)$ or not $O(3)$~?}

Master fields $\phi_c (x)$ for baryon number violation
which have been suggested in the literature
(e.g.-- the \iibar pair, the valley, $I^*$)
have typically been chosen because they approximately
(or in the $I^*$ case, exactly) extremize the action $S$.
However, solutions which extremize $S$ typically have {\it more}
symmetry than those that extremize $S_{eff}$.
As we have shown,
one can only strictly expect at most $O(2) \times T \times Z$ symmetry
of $\phi_c (x)$, while configurations that extremize (or nearly extremize)
$S$ typically have more symmetry
(the examples mentioned above are all $O(3)$ symmetric).
One might expect that at energies of order $E_{sph} \sim M_W / g^2$,
where the initial state factor is important,
the solutions which extremize $S_{eff}$ will cease to exhibit
the full $O(3)$ symmetry.
This could happen suddenly, at some critical energy $E_c$,
or gradually, with only the zero energy master field for $S_{eff}$
exhibiting exact $O(3)$ symmetry.

Here we show that in the case of on-shell scattering there is
a dynamical reason for $\phi_c (x)$ to be $O(3)$ symmetric.
Without loss of generality,
the master field can be expressed in terms of a finite number $N$
of arbitary functions $\phi_i (x) $,
each ``centered'' at $x=x_i$,
\Eq
\phi_c (x) = \sum^N_{i=1} ~ \phi_i ( x - x_i)~.
\Endl{sum}
We will first demonstrate the asymptotic $O(3)$ symmetry of $\phi_c$
by showing that quite generically:
\begin{enumerate}
\item $\phi_i(x)$ are asymptotically O(4) symmetric for all $i$, and
\item $x_i$ are collinear for all $i$,
\end{enumerate}
when $\phi_c$ is the saddlepoint of an on-shell scattering amplitude.

Recalling the scattering amplitude (\ref{sca}),
the terms which arise in the initial state factor
\Eq
\Fp (k_1) ~\Fp(k_2) ~\Fp^* (k_1) ~\Fp^* (k_2) ~,
\Endl{IS}
can each be written as
\Eq
\Fp (k) = \sum^N_{i=1}~ \Fp_i (k)~ e^{i k \cdot x_i} ~.
\Endl{FTsum}
This yields an initial state factor with
all possible terms of the form
\Eq
\sum_{abcd}^N~
e^{ik_1 \cdot x_a} e^{ik_2 \cdot x_b}
e^{- ik_3 \cdot x_c} e^{-ik_4 \cdot x_d}
{}~\Fp_a (k_1) ~\ldots~ \Fp^*_d (k_2) ~,
\Endl{terms}
each leading to a different extremization problem.

On-shell scattering requires a pole in at least some
of the $\Fp_i (k)$ at $k_{\mu} k^{\mu} = m^2$
(or, in Euclidean space, $(k^E_{\mu})^2 = - m^2$),
where $m^2$ is the mass of the physical particle.
In other words,
\Eq
\Fp_i (k) \: = \: { A(k^{\mu}) ~\over~ k_{\mu}^2 - m^2 } ~+~ \ldots ~,
\Endl{pole}
where the ellipsis denote nonsingular terms and $A(k^{\mu})$ is regular.
In any frame, we have  $k_{\mu}^2 = E^2 - \vec{k}^2$.
Since the pole in $\Fp_i (k) $ is determined by the
asymptotic behavior ($|x| \rightarrow \infty$) of the function $\phi_i (x)$,
we find that at large distances $\phi_i (x)$ must posess $O(3,1)$ symmetry
in Minkowski space and $O(4)$ symmetry in Euclidean space~\cite{BV}.
This occurs naturally when initial state factors are neglected
and $\phi_c$ is chosen to be a saddlepoint of the action.
Then, \mbox{$\phi_c$ solves} the free massive wave equation at large $|x|$,
which in turn implies an on-shell pole~\footnote{See Espinosa in \cite{BV}}.
Of course, in the problem at hand, we are concerned with
the effect of the initial state factors, or  a saddlepoint of
the {\it effective} action (\ref{seff}).
Such configurations may not in general solve the free wave equation at
large $|x|$.

Henceforth,
we will suppress any terms in the sum (\ref{FTsum}) which do not
have the proper asymptotic behavior.
Such terms are irrelevant to the initial state factor,
and their local effects can be included in a redefinition of the
$\phi_i (x)$ terms.

Now consider the phase factors in (\ref{terms}).
For example, when $a=b$ and $c=d$, we obtain
\Eq
e^{i (k_1 +k_2) \cdot (x_a - x_c)} ~=~ e^{i\, 2E \, (t_a - t_c)} ~,
\End
for the exponential factor.
In Euclidean space,
and for energies $E \sim 1/g^2$,
this term has the familiar effect of making the saddlepoint
of (\ref{fsa}) energy dependent and
is responsible for the growth of the total cross section at low
energy~\cite{BV}.
Careful examination of the other terms appearing in (\ref{terms})
reveals that they always exhibit symmetry about a {\it single} axis,
or are subleading in their effect.
The leading terms therefore always lead to $O(3)$ symmetric configurations,
at least at the asymptotic level.

Let us state more precisely what we mean by the asymptotic behavior
of $\phi_i(x)$.
Without loss of generality, we can write $\phi_i (x)$ as
\Eq
\phi_i (x)~ = ~ f ( x^{\mu} ) ~ e^{- m |x|},
\Endl{feq}
where we have suppressed the Lorentz (in the case of $\phi = A^{\mu}$)
and gauge indices of $f ( x^{\mu} )$.
Now, the Fourier transform is defined as
\Eq
\Fp_i (k) = \int d^4 x~ e^{i k \cdot x} ~ \phi_i (x) ~.
\Endl{FT}
For $\phi_i (x)$ vanishing exponentially as $|x| \rightarrow \infty$,
$\Fp_i (k)$ is analytic for real
values of $k$, and only has
singularities where the exponential falloff in $\phi_i (x)$
is cancelled by the $e^{i k \cdot x}$ factor. In order for the amplitude to
describe on-shell scattering, this
means imaginary momenta $(k^E_{\mu})^2 = - m^2$.
Immediately, this requires that $f ( x^{\mu} )$ not have
exponential behavior as $|x| \rightarrow \infty$ (i.e. - $| f ( x^{\mu} ) |
\sim |x|^n$ or $\log ~|x|$).
Also, it is clear that in order that (\ref{FT}) diverge as
$k^2 \rightarrow - m^2$, $|f ( x^{\mu} )|$ must not fall off too fast at large
$x$.

More specifically, the precise form of $A(k)$ can be determined from
$f ( x^{\mu} )$. First write
\Eq
A(k) ~=~
\lim_{k^2 \rightarrow  - m^2} ~ (k^2 + m^2)~ \int d^4x~
e^{i k \cdot x}\, e^{- m|x|}\, f(x^{\mu}) ~.
\End
Now let $k^{\mu} = k_0^{\mu} + \epsilon^{\mu}$,
where $k_0^{\mu}$ is a fixed four-vector satisfying $k_0^2 = - m^2$.
We choose $k^{\mu}$ and $\epsilon^{\mu}$ antiparallel so that
$|k_0 +\epsilon| = m - \epsilon$,
where $\epsilon \equiv |\epsilon^{\mu} |$.
Then we have
\if l\format
\Eqa
\lefteqn{A(k) ~=~} & & \\
& & \lim_{\epsilon \rightarrow 0} ~  2i m \epsilon
\int_0^\infty dr\, r^3\, d \Omega\int_{-1}^{+1} du~\sqrt{1 - u^2}~
e^{ (m - \epsilon) r u}\, e^{- mr} \, f( x^{\mu}) ~. \nonumber
\Enda
\else
\Eq
A(k) ~=~ \lim_{\epsilon \rightarrow 0} ~  2i m \epsilon
\int_0^\infty dr\, r^3\, d \Omega\int_{-1}^{+1} du~\sqrt{1 - u^2}~
e^{ (m - \epsilon) r u}\, e^{- mr} \, f( x^{\mu}) ~.
\End
\fi
Here, we have defined $r = |x|$,
$u = \cos~ \theta$ as the cosine of the angle
between $x^{\mu}$ and $k^{\mu}$,
and $\Omega$ as the remaining angular variables.

Now consider the $u$ integral. Only the part of the integral
with $u \simeq 1$ will contribute to the residue
$A(k)$, since for $u \neq 1$,
the $r$ integral is convergent.
In light of this observation, we split the $u$ integral into two parts,
\Eq
\int_{-1}^{+1} du ~\ldots~ = ~ \int_{-1}^{1- \delta} du ~\ldots~
+ ~ \int_{1 - \delta}^{1} du ~\ldots ~.
\End
Only the second part can contribute to $A(k)$.
In the region $u \simeq 1$,
we expand the product
$\sqrt{ 1 - u^2}\, f(u,r,\Omega) = \sum_n v^n f_n (r, \Omega)$, where
$v \equiv 1 - u$.
The $f_n$ are essentially determined by derivatives
$\partial^n  f / \partial u^n$ evaluated at $u = 1$.
There may be fractional
powers of $v$ in the expansion
(in particular, $\sqrt{2\, v}$ coming from $\sqrt{1-u^2}\,$),
however for simplicity we use the index $n$.

Each term in the $v$ expansion leads to an integral
\if l\format
\Eqa
\lefteqn{
e^{- \epsilon r}~ f_n (r, \Omega)~\int_0^{\delta} dv~ v^n
e^{- (m - \epsilon) rv}   ~=~} & & \label{fsubn} \\
& & \hspace{4.0cm}
e^{- \epsilon r}~
{f_n (r, \Omega) \over (m\, r)^{n+1}}
{}~\int_0^{\delta'}
dx~ x^n~e^{-x} ~. \nonumber
\Enda
\else
\Eq
e^{- \epsilon r}~ f_n (r, \Omega)~\int_0^{\delta} dv~ v^n
e^{- (m - \epsilon) rv}   ~=~
e^{- \epsilon r}~
{f_n (r, \Omega) \over (m\, r)^{n+1}}
{}~\int_0^{\delta^\prime}
dx~ x^n~e^{-x} ~.
\Endl{fsubn}
\fi
Substituting this into the $r$ integral yields terms like
\Eq
\int dr ~r^{-n+2}  ~  f_n (r, \Omega) ~  e^{- \epsilon r} ~,
\End
which can only yield $1/ \epsilon$ poles when the large $r$ behavior of $f_n$
exactly cancels that of $r^{-n+2}$.
Therefore, terms with $f_n \sim r^l$ contribute a finite amount to $A(k)$ when
$l = n-2$. Those with
$l < n-2$ do not affect $A(k)$, while those with $l > n-2$ give an infinite
contribution to $A(k)$ and must
be excluded.

As an application of the above results,
let us compute the residue associated with a deformed,
non-$O(4)$ symmetric configuration $\phi_i $.
Without loss of generality,
the configuration may be expressed as a function with
$O(3)$ symmetry about the $x_3$-axis
\Eq
\phi_i(x) \: = \: h(r^2, x_3^2) \: {K_1(m\, r)\over m\, r}  ~.
\Endl{axiston}
The second factor here  is a convenient
combination, for it satisfies the free massive wave equation.
One recognizes immediately the exponential
decay of the massive field at large distances,
from the asymptotic expansion
$K_1(z)\sim e^{-z}/\sqrt{2z/\pi}$ as $|z|\rightarrow\infty$.
Although we consider only the case of a scalar field here,
the case of massive vector fields entails only the additional
complication of indices.

Now we can envision at least two cases for the behavior of $f$.
If the deformation from $O(4)$ symmetry
is entirely contained within a finite volume,
then the on-shell residue is unchanged.
This is the case if $f$ reduces to a constant or falls off
at large (Euclidean) distances.
A less trivial case occurs when the deformation persists to
large distances.
We choose the following $O(3)$ symmetric
function as an example to illustrate the idea:
\Eq
h^{(\alpha)}(r^2, x_3^2) \: = \:
1  \: + \: \alpha^2\, x_3^2/r^2 ~,
\Endl{halpha}
where $0<\alpha^2<\infty$ parametrizes the deformation from
$O(4)$ symmetry.

At large $|x|$, the configuration (\ref{axiston}) resembles
(\ref{feq}) with
\Eq
f^{(\alpha)}(r^2, x_3^2) \: = \:
\sqrt{{\pi\over 2}}~{1 \over (mr)^{3/2}}~\left(\, 1  \: + \:
\alpha^2\, x_3^2/r^2\,\right)  ~.
\Endl{whatever}
Then, a direct application of (\ref{fsubn}) yields a term with $n = 1/2$
\Eq
A(k)~\rule[-3mm]{0.2mm}{8mm}_{\: k = \pm i m} \: \sim \:
\lim_{\epsilon \rightarrow 0} ~  2i m \epsilon~\left( 1 + \alpha^2 \right)
\int_0^\infty dr~e^{-\epsilon r} ~,
\End
plus terms with $n > 1/2$ which vanish as $\epsilon\rightarrow 0$.

Our choice of example (\ref{halpha}) allows us to check this
result with an exact calculation of the residue.
After integrating over all angles, $\Omega, u$, in the Fourier
transform, we are left with the following  integral over $r$:
\Eq
\Fp^{(\alpha)}(k) \: = \: {4\pi^2\over k} \,
\left(1 + \alpha^2\right)
\,\int_0^\infty dr\, r \: K_1(mr)\, J_1(kr)  ~,
\End
up to terms $\propto J_{n>1}$ which do not contribute to
the on-shell pole.
Since $J_n(z)\sim e^{iz}/\sqrt{z}$ as $|z|\rightarrow\infty $,
the integrand has a pole when \mbox{$ik = \pm~m$.}
The residue of this pole is
\Eq
A(k)~\rule[-3mm]{0.2mm}{8mm}_{\: k = \pm i m} \: = \:
{4\pi^2 \over m}\,\left( 1 + \alpha^2 \right) ~.
\End

As we can see from the above analysis, the dependence of $A(k)$ on parameters
specifying
$f ( x^{\mu} )$ is
weaker than exponential. Changing $f ( x^{\mu} )$ slightly
(say from $O(4)$ to $O(3)$ symmetric),
while keeping the correct pole structure,
will therefore only affect the initial state factors in (\ref{fsa})
a subleading way.
On the other hand,
since most of the action density of the configuration $\phi_c (x)$ is
in the region controlled by $f ( x^{\mu} )$,
a small change in $f ( x^{\mu} )$ will tend to change the action of
the configuration,
which occurs in the exponent with prefactor $1/g^2$.
In general, since $O(3)$ configurations tend to have lower
actions than less symmetric configurations, the action should increase.
Unless there are factors of $e^{1/g^2}$ in the residue $A(k)$,
the increase in action cannot be compensated by
an increase in the initial state factor.
However, our analysis implies that
factors of $e^{1/g^2}$ in $A(k)$ can only result from
an exponentially large magnitude or derivative of the function
$f(x^{\mu})$,
which in turn leads to an exponentially large action.
This implies that
the amplitude (\ref{fsa}) is likely to always be dominated by fully
(not just asymptotically) $O(3)$ symmetric configurations !
In this case,
the effective action for an on-shell scattering amplitude
reduces from (\ref{seff}) to the familiar low energy ($E\ll E_{sph}$)
form at order $\sim 1/g^2$,
\Eq
S_{eff}\left[\,\phi\,\right]\: = \: S\left[\,\phi\,\right]  \, - \,
E\, R ~,
\Endl{seffred}
where $R$ is the Euclidean separation between two $\phi_i$ in (\ref{sum}).
It would seem that non--$O(3)$ symmetric configurations are irrelevant
at the level of the semiclassical exponent.

\mysection{Fermion Scattering}

Spacetime symmetries have similar implications for
amplitudes involving fermion fields.
The forward scattering amplitude involving four fermion fields
is derived from a correlator
\if l\format
\Eqa
\lefteqn{
\langle ~\Psi (k_1) ~\Psi(k_2) ~\Psi^* (k_1) ~\Psi^* (k_2) ~\rangle
\: = \: } & & \label{fermionfsa}\\
& & \hspace{1.5cm} N \int D\phi~DA~D\Psi~D\bar{\Psi}
{}~~\Psi (k_1) \ldots \Psi^* (k_2) ~e^{i S\left[\Phi,A,\Psi\right]} ~,
\nonumber
\Enda
\else
\Eq
\langle ~\Psi (k_1) ~\Psi(k_2) ~\Psi^* (k_1) ~\Psi^* (k_2) ~\rangle
\: = \: N \int D\phi~DA~D\Psi~D\bar{\Psi}
{}~~\Psi (k_1) \ldots \Psi^* (k_2) ~e^{i S\left[\Phi,A,\Psi\right]} ~,
\Endl{fermionfsa}
\fi
where Lorentz indices and isopin indices have been suppressed.
Again assuming the existence of a semiclassical approximation,
the fermion fields are to be expanded, $\Psi = \Psi_c + \delta\Psi$
around the classical solutions to the (Euclidean) Dirac equation,
in the classical gauge and scalar background,
\Eq
\left(\, i\!\not\!\! D_c \: + \: y \Phi_c \,\right)~\Psi_c
\: = \: 0  ~.
\Endl{dirac}
Here the Yukawa coupling to the Higgs scalar field is $y$
and the (Euclidean) gauge-covariant derivative is
$D_c^\mu = \partial^\mu + \AMC$.
The bosonic background fields $\{\Phi_c, \AMC\}$
are determined by the same saddlepoint calculation described in
the previous sections.
The fermion determinant is not exponential and does not affect
the bosonic equations of motion at leading semiclassical order.

At low energies, where the approximate saddlepoint background
is a widely separated \iibar pair,
an index theorem ensures  the existence of
approximate fermion zero mode solutions to (\ref{dirac}).
Their contribution to the amplitude realizes a projection onto
fermion number violating intermediate states in
(\ref{fsa}) or (\ref{fermionfsa})
and the identification of an unambiguous non-perturbative
contribution to the total cross-section.
However,
several authors (see latter references in ~\cite{ZEROMODES})
have demonstrated that these zero modes disappear from the spectrum
at sufficiently small \iibar separation,
as the topological charge in a finite volume decreases from $\pm 1$
when the separation decreases.
The index theorem suggests that we must require that the bosonic
background must have topological charge $+1$ and $-1$
in finite volumes separated by some distance in Euclidean time
in order to ensure baryon number violation.

A solution of (\ref{dirac}) has {\it at most} the symmetries of
the bosonic background $\{\Phi_c, \AMC\}$.
Arguments similar to those in Section 2 require that,
in the CM frame,
either $\Psi_c$ is $O(2) \times T \times Z$ invariant,
or there are a family of $\Psi_c$
related by these symmetries which contribute equally to the amplitude.
Then,
the bosonic saddlepoint fields $\{\Phi_c, \AMC\}$ must also have
at least these symmetries.
For the case of on-shell scattering,
we again must require that $\Psi_c$ have asymptotic $O(3)$ symmetry,
as described in Section 3.
Then,
the bosonic fields must also be asymptotically at least $O(3)$ symmetric.
We arrive at conclusions for the bosonic fields which are
identical to the case of bosonic scattering (\ref{fsa}),
even though the bosonic fields do not appear in the initial state factor
of (\ref{fermionfsa}).
The dynamics of the fermion-boson interactions impose these restrictions
on the bosonic fields indirectly.

\mysection{Conclusions}
We have found that spacetime symmetries yield nontrivial constraints
on saddlepoints that dominate forward scattering amplitudes.
These results apply in the case of fermionic as well as bosonic scattering.
We find that in the case of on-shell scattering Lorentz invariance requires
$O(3)$ asymptotic behavior for the saddlepoint.
We argued that dynamical considerations extend this symmetry to
the entire saddlepoint configuration,
at least as far as the semiclassical exponent is concerned.
Therefore,
the saddlepoint calculation which has become standard at low energies
($E\ll E_{sph}$),
namely the extremization of an effective action of the form (\ref{seffred}),
should remain qualitatively unchanged even at high energies
($E\simeq E_{sph}$.)
We hope that these conclusions will simplify the problem of extremizing
scattering amplitudes derived from correlators such as (\ref{fsa}) or
(\ref{fermionfsa}).

\vskip 0.5cm
\centerline{\bf Acknowledgements}
\vskip 0.1in
The authors would like to thank F. Klinkhamer for useful discussions.
The authors are grateful to the participants of the Aspen Center for
Physics workshop on Baryon Number Violation, and especially to the
organizers, Larry McLerran and Valery Rubakov,
for providing a stimulating environment where this work was begun.
TMG acknowledges the support of the National Science
Foundation under grant \mbox{NSF-PHY-90-9619} and the
state of Texas under grant \mbox{TNRLC-RGFY-93-292.}
SDH acknowledges support from the National Science Foundation
under grant \mbox{NSF-PHY-87-14654,}
the state of Texas under grant \mbox{TNRLC-RGFY-106,}
the Harvard Society of Fellows and an SSC Fellowship.
The authors also acknowledge the support of the
White Horse and Railroad Foundation.

\if l\format
\newpage
\fi

\end{document}